\documentclass[preprint,aps,tightenlines,floatfix,showpacs]{revtex4}

\usepackage{graphics}
\usepackage{bm}
\usepackage{amsmath}
\usepackage{amssymb}

\begin{document}

\title{Properties of  effective interactions and the excitation of
6$^-$ states in ${}^{28}$Si}

\author{Y. J. Kim$^{a}$}
\email{yjkim@cheju.ac.kr}
\author{K. Amos$^{a}$} 
\email{amos@physics.unimelb.edu.au}
\author{S. Karataglidis$^{a,b}$}
\email{kara@physics.unimelb.edu.au}
\affiliation{${}^{a}$ School of
Physics, The University of Melbourne, Victoria 3010, Australia}
\affiliation{${}^{b}$
Department of Physics and Electronics, Rhodes University,
Grahamstown 6140, South Africa}

\date{\today}

\begin{abstract}
Cross-section and analyzing power data from $(p,p^\prime)$ scattering to 
the $6^-$ states at 11.58 and 14.35 MeV in ${}^{28}$Si, taken with
energies of 80, 100, 134, and 180 MeV protons, have been analyzed using a 
distorted wave approximation with microscopically defined wave functions.
The results, taken in conjunction with an analysis of an M6 electron 
scattering form factor, suggest that the two states exhaust respectively,
$\sim 50$\% and $\sim 60$\% of the strengths of isoscalar and isovector 
particle-hole excitations from the ground state. The energy variation of data 
also suggests that the non-central components of the effective interactions
at 80 and 100 MeV may need to be enhanced.
\end{abstract} 

\maketitle

\section{Introduction}

In a recent publication~\cite{Am05}, we demonstrated that analyses of 
complementary reaction data, from the scattering of electrons, of pions,
and of protons leading to (dominantly isoscalar and isovector) $4^-$
states in ${}^{16}$O in the vicinity of 19 MeV excitation, could ascertain
the degree of isospin and configuration mixing in the descriptions of
those unnatural parity states.
The existence of data from three distinct $4^-$ excitations in ${}^{16}$O 
bespoke
of configuration mixing, given that a pure stretched scheme could only
yield two such states.  Those results also validated the tensor and
spin-orbit character of the two-nucleon $NN$ effective interaction~\cite{Am00}
used in the distorted wave approximation (DWA) analyses of proton scattering
(at 135 MeV) since those terms
dominate scattering amplitudes of M4 transitions. Data were available
from proton scattering (inelastic and charge exchange) and the
characteristics (magnitude and shape of cross sections and spin observables
as well as of form factors) clearly formed the conclusions drawn.

There are other nuclei with states that, in the simplest structure concept,
are stretched particle-hole excitations from the ground. To our
knowledge however, none have the variety of measured data as with the 
$4^-$ excitations in ${}^{16}$O. One such case is the excitation of 
$6^-$ states in ${}^{28}$Si at 11.58 and 14.35 MeV above the ground. They
are of interest to us first as proton scattering data exciting them have been 
measured 
for incident energies of 80, 100, 134, and 180 MeV and of both differential
cross sections and analyzing powers~\cite{Ol84}.
We consider this set of data herein in part to see if such 
suffice to identify configuration and/or isospin mixing within the actual 
states. The spectrum of ${}^{28}$Si already infers that the particle-hole
strength for excitation of the $6^-$ states could be fragmented. 
Pure stretched states formed just by promotion of a $0d_{\frac{5}{2}}$
particle into the $0f_{\frac{7}{2}}$ orbit, would result in just two 
transitions; the 
isoscalar and isovector combinations of the proton and neutron excitations.
But, in the excitation energy range from 11 to 15 MeV, there are five known 
levels of that spin-parity~\cite{En98}.

Added interest lies in the fact that Skyrme-Hartree-Fock (SHF) calculations 
of the ground state of ${}^{28}$Si have been made recently~\cite{Ri03} with  
the resulting wave functions giving form factors in good agreement with 
available data from electron scattering. Two forms of the 
Skyrme interaction were used, of which we consider only the SHF wave functions
determined using the so-called SkX$_{csb}$ interaction~\cite{Br00}. 
The SkX$_{csb}$ Hamiltonian is based  on the SkX
Hamiltonian~\cite{Br98} with a charge-symmetry-breaking (CSB) interaction
added to  account  for  nuclear displacement energies.  Generally,
with this SHF method, good agreement between theory and experiment   has   been
achieved   in    extensive   comparisons of  measured  nuclear   charge-density 
distributions with calculated values for $p$-shell,  $sd$-shell, and $pf$-shell
nuclei and also with some  selected  magic and semi-magic nuclei up to 
${}^{208}$Pb. 
They have been used more recently~\cite{Am06} in 
analyses of total reaction cross sections of protons
by which a new test of the neutron matter distribution was shown to be feasible.
While much success has been had using the SHF densities and wave functions
generated using the SkX$_{csb}$ model, the canonical wave functions may not 
have a desirable long range character. So we have also used Woods-Saxon (WS)
single nucleon bound state wave functions in place of them for comparison
of scattering results. 

Finally, as data have been taken at a set of energies, and as such unnatural 
parity  transitions are dominated by the non-central components of the 
transition operator~\cite{Am05}, these observables may probe the tensorial
character of the effective $NN$ interactions sensitively. That is so as the 
(complex) effective $NN$ interactions~\cite{Am00} are energy as well as density
dependent.  Furthermore, the energy region is one for which past $g$-folding
studies~\cite{Am00} have successfully predicted elastic scattering 
observables. It is also one for which DWA calculations, made using the distorted
wave functions from those $g$-folding optical potentials with the effective
$NN$ $g$-matrix as the transition operator, gave good predictions of
inelastic scattering observables to states of nuclei for which the spectroscopy
was well determined~\cite{Am00}. An example is of excitation of the 
$2^+_1$ (4.43) MeV excitation in ${}^{12}$C
when described by a complete $(0+2)\hbar\omega$ shell model study.
But those successes, primarily, are measures of the central force components
of the effective $NN$ interactions.

In the following section we describe the structure model for ${}^{28}$Si and 
discuss the excitations to the 6$^-$ states.
That structure is tested by comparing predictions of the transverse magnetic
form factor with data from electron scattering. Then, in Sec.~\ref{DWAelements},
we present relevant details of the DWA excitation matrix elements. The results 
of our DWA
calculations are presented thereafter in Sec.~\ref{Results} and conclusions are 
given in Sec.~\ref{Conclusions}.

\section{The structure models}
\label{Structure}

As noted above, we have used  SHF and WS wave functions to describe the
nucleon orbits in ${}^{28}$Si. The SHF calculation~\cite{Ri03} considered
was that built with the SkX$_{csb}$ interaction. Consequently proton and 
neutron 
densities differ slightly; how much being  shown below in Fig.~\ref{Fig1}.
In those studies, the single-particle occupancies $n_j$ were constrained to 
values 
obtained from shell-model calculations. That constraint resulted in a 
significantly improved agreement with most electron scattering form factors
considered in Ref.~\cite{Ri03}. For ${}^{28}$Si, these
occupancies are listed for the dominant orbits in Table~\ref{Occup}.
\begin{table}[h]
\caption{ \label{Occup}
The shell occupancies and binding energies taken for the ground
state of ${}^{28}$Si.}
\begin{ruledtabular}
\begin{tabular}{cccccc}
$\varphi_j$ & Occupancy & B.E. (MeV) & $\varphi_j$ & Occupancy & B.E. (MeV)\\
\hline
$0s_{\frac{1}{2}}$ & 2.0 & 29.7 & $0d_{\frac{5}{2}}$ & 4.623 & 2.51\\
$0p_{\frac{3}{2}}$ & 4.0 & 16.1 & $1s_{\frac{1}{2}}$ & 0.704 & 0.7\\
$0p_{\frac{1}{2}}$ & 2.0 & 12.4 & $0d_{\frac{3}{2}}$ & 0.673 & 0.5\\
\end{tabular}
\end{ruledtabular}
\end{table}
\begin{figure}[h]
\scalebox{0.7}{\includegraphics*{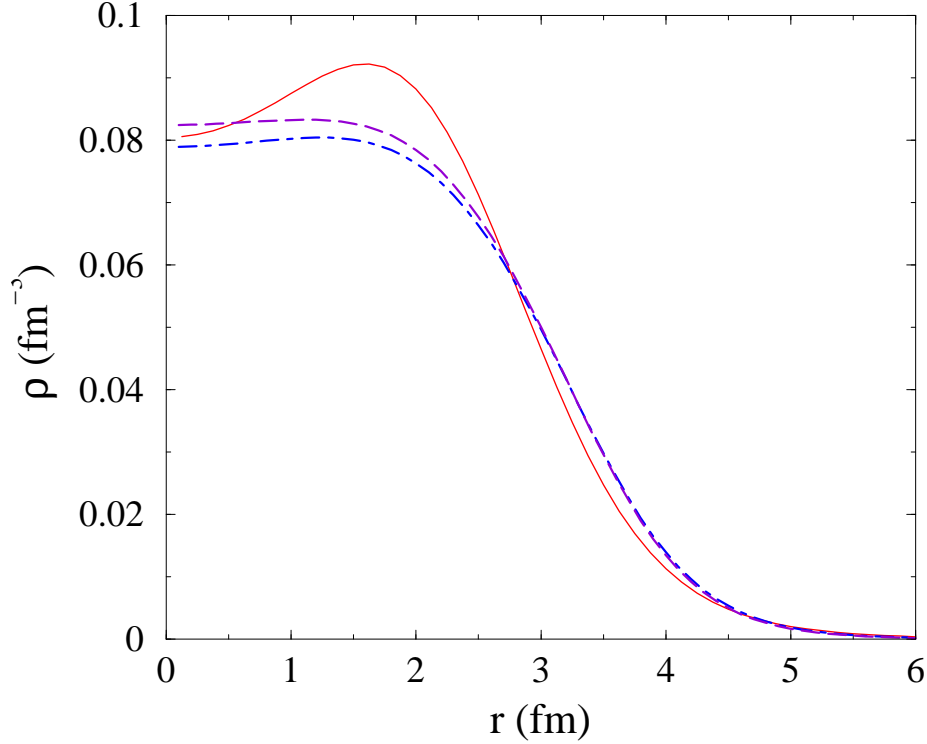}}
\caption{\label{Fig1}(Color online)
The SHF model densities, proton (dot-dashed curve) and neutron
(dashed curve) for ${}^{28}$Si are compared to that found using
WS wave functions but with the same shell occupancies. The proton and 
neutron WS distributions are identical with form displayed by the solid curve.}
\end{figure}
The binding energies listed in Table~\ref{Occup} were used with a Woods-Saxon
bound state potential (radius $r_0 = 1.2$ fm, diffuseness $a = 0.6$ fm) 
to define the WS bound state functions which form an alternative set to those 
given by the SHF calculation.

The SHF model densities (proton and neutron) are compared in Fig.~\ref{Fig1}
to the WS proton (neutron) density of the ground state of ${}^{28}$Si. 
The difference between the SHF (proton and neutron) matter densities 
and the WS one with its surface peak are evident. The surface
peaking behavior of the WS distribution tends more like that
Richter and Brown~\cite{Ri03} extracted from the electron form factor than
does their SHF result.

\subsection{One-body density matrix elements}

One-body density matrix elements (OBDME) are required both in forming the 
optical potentials, with which we predict elastic scattering observables,
and in the DWA amplitudes with which we predict inelastic and charge 
exchange observables.  With a particle-hole matrix element cast in 
irreducible tensor form by 
\begin{align}
&\left\langle \Psi_{J_fM_f} \left|  a^\dagger_{j_2 m_2}\ a_{j_1 m_1}
\right| \Psi_{J_iM_i} \right\rangle =
\nonumber\\
&\hspace*{1.0cm}
\sum_{I(N)} (-1)^{(j_1-m_1)} 
\left\langle j_1\,j_2\,m_1\,-m_2 \left. \right| I\, -N \right\rangle
\ \left\langle \Psi_{J_f M_f} \left| 
\left[a^\dagger_{j_2} \otimes  {\tilde a}_{j_1}\right]^{IN}
\right| \Psi_{J_i M_i} \right\rangle\ ,
\end{align}
use of the Wigner--Eckart theorem gives
\begin{align}
&\left\langle \Psi_{J_fM_f} \left|  a^\dagger_{j_2m_2}\ a_{j_1m_1}
\right| \Psi_{J_iM_i} \right\rangle = 
\nonumber\\
&\hspace*{1.0cm}\sum_{I(N)}
 (-1)^{(j_1-m_1)} \left\langle
j_1\,j_2\,m_1\,-m_2\left. \right| I\, -N
\right\rangle
\times \frac{1}{\sqrt{2J_f+1}}
\left. \left\langle J_i\,I\,M_i\,N \right| J_f\,M_f \right\rangle
\ S_{j_1 j_2 I}\ , 
\end{align}
where the OBDME is
\begin{equation}
S_{j_1 j_2 I} = \left\langle \Psi_{J_f M_f} \left|\left|
\left[ a_{j_2}^\dagger \otimes {\tilde a}_{j_1} \right]^{I}
\right|\right| \Psi_{J_i M_i} \right\rangle\ .
\end{equation}

For elastic scattering from ${}^{28}$Si (a spin-zero target), the 
OBDME are the ground state expectations of the particle-hole operator
for zero angular momentum transfer ($I = 0$). Often those numbers reduce
simply to being the fractional shell occupancies of nucleons in the ground
state $\sigma_j = n_j/(2j + 1)$.

For the excitation of the stretched $6^-;T$ states, transition OBDME
for $I = 6$ are required and similar to the development~\cite{Am05} for the 
stretched $4^-;T$ states in ${}^{16}$O, in the case of ${}^{28}$Si,
with states given by
\begin{equation}
\left| 6^-; T \right\rangle = 
\left| \left( d^{-1}_{\frac{5}{2}} f_{\frac{7}{2}} \right) 6^-; T\right\rangle
= \frac{1}
{\sqrt{\left[\left(1-\sigma_{\frac{7}{2}}\right)\sigma_{\frac{5}{2}} \right]}}
\ \left[ a^\dagger_{\frac{7}{2},\frac{1}{2}} \otimes 
{\tilde a}_{\frac{5}{2}, \frac{1}{2}} 
\right]^{6, T}_{M_6, M_T = 0} 
\left| 0^+; 0\right\rangle\ ,
\end{equation}
where $\sigma_{j}$ is the fractional occupancy of protons (and of neutrons)
in the shell $j$ in the ground state (listed in Table~\ref{Occup}), the OBDME 
are
\begin{equation}
S_{\frac{5}{2} \frac{7}{2} 6}(x) 
= \sqrt{\frac{13}{2}}
\left\{\delta_{T,0} + \delta_{T,1} (-1)^{\left(1/2 -x\right)}\right\} 
\sqrt{\left[\left(1-\sigma_{\frac{7}{2}}\right)\sigma_{\frac{5}{2}} \right]}\ ,
\end{equation}
when $x = \pm \frac{1}{2}$ for a neutron and a proton respectively.
With the shell occupancies listed in Table~\ref{Occup}, the OBDME for 
excitation of stretched $6^-$ states has the magnitude of 2.238.

\subsection{Electron scattering and the M6 transverse form factor}

Electron scattering exciting the isovector $6^-$ state at 14.35 MeV
has been measured~\cite{Ye80} and the M6 form factor extracted in the
range of momentum transfer from 1 to nearly 3 fm$^{-1}$. That form factor 
was extracted from cross section data at two scattering angles using
the plane wave definition,
\begin{equation}
\frac{d\sigma}{d\Omega} = 4\pi \sigma_M R
\left[F_{C\lambda}^2(q) + \left\{ \frac{1}{2} + \tan^2(\theta/2)\right\}
F_{T\lambda}^2(q) \right]\ .
\label{eescat}
\end{equation}
\begin{figure}[h]
\scalebox{0.7}{\includegraphics*{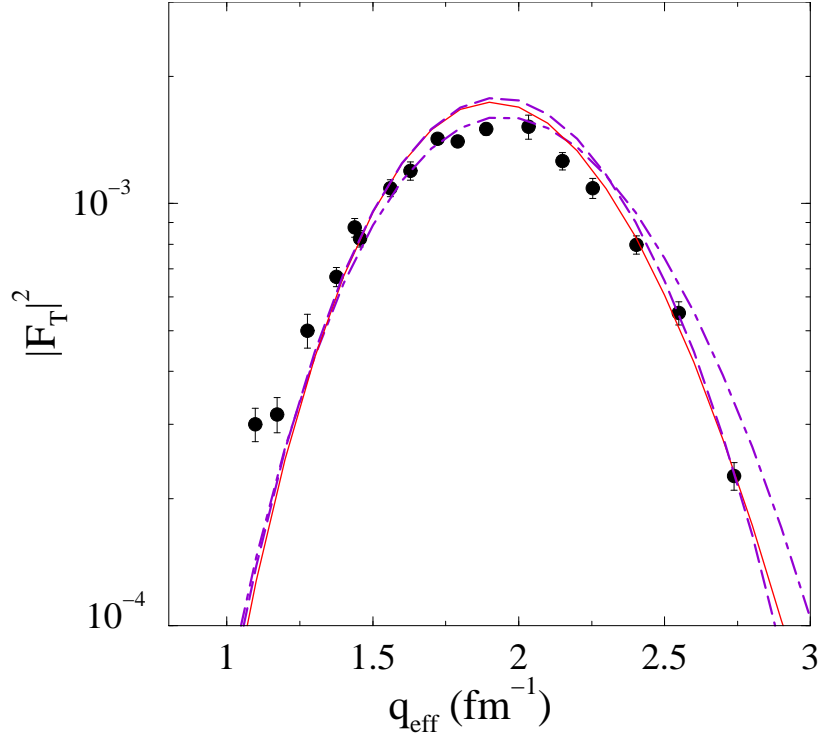}}
\caption{\label{Fig2}(Color online)
The M6 transverse form factors calculated using oscillator (dashed curve),
SHF model (solid curve), and WS (dot-dashed curve) wave functions compared 
with data~\cite{Ye80}.}
\end{figure}
Excitation of the $6^-; T=1$ state yielded $\left|F_T\right|^2 = F_{T6}^2(q)$.
The other factors in Eq.~(\ref{eescat}) are the Mott scattering 
cross section $\sigma_M$ and a recoil factor $R$~\cite{Ye83}. The M6 
form factor is shown in Fig.~\ref{Fig2} where it is
plotted against an effective momentum transfer $q_{\text{eff}}$, 
to take into account the effect of acceleration of the electron by the 
Coulomb field of the target~\cite{Ye83}.  In this figure,
the data are compared with three calculated results. That depicted by
the dashed curve was found using  $0\hbar\omega$ shell model (oscillator
length $b = 1.74$ fm)
wave functions while that depicted by the solid curve was obtained using the 
SHF wave functions. The third, dot-dashed, curve is the result found 
by using WS wave functions.
All calculated results determined by using the pure
particle-hole description of the excited state (with shell occupancies given 
in Table~\ref{Occup}, and so the spectroscopic amplitude of 2.238)
have been scaled by 0.36. The results 
suggest that the strength of this particle-hole state is dissipated among 
others, with the 14.35 MeV state accounting for $\sim$60\%.

\section{Elements of the DWA calculations}
\label{DWAelements}

 Cross sections for inelastic proton scattering exciting the $6^-$ 
states in ${}^{28}$Si have  been  evaluated  using  a fully 
microscopic DWA theory of the processes~\cite{Am00}.  As all details 
have been given in that review, only salient features are 
dealt with in this section.  The distorted
waves are generated from optical potentials formed  by folding  an
effective in-medium $NN$ interaction 
with the OBDME of the target states. 
At each energy, the effective $NN$ interaction is generated from a mapping
to solutions of Brueckner-Bethe-Goldstone equations (the $NN$
$g$ matrices). 
In coordinate space that effective $NN$ interaction is a mix 
of central, two-body spin-orbit and tensor forces all having  form 
factors that are sums of Yukawa functions.     Then with the Pauli
principle taken into account,   optical potentials from the folding
are complex, nonlocal, and energy dependent.   Such are formed and
used in the DWBA98 program~\cite{Ra98} to predict elastic 
scattering observables. That same program finds distorted wave functions 
from those potentials for use in  DWA  calculations  of  the inelastic 
scattering  cross sections and spin observables.   
The transition amplitudes for 
nucleon inelastic scattering  from a nuclear target 
have the form~\cite{Am00}
\begin{eqnarray}
{\cal T} &=& T^{M_fM_i\nu^\prime\nu}_{J_fJ_i}(\Omega_{sc})
\nonumber\\
&=&\left\langle \chi^{(-)}_{\nu^\prime}({\bf k}_o0)\right|
\left\langle\Psi_{J_fM_f}(1\cdots A) \right|
\; A{\bf g}_{\text{eff}}(0,1)\;
 {\cal A}_{01} \left\{ \left| \chi^{(+)}_\nu ({\bf
k}_i0) \right\rangle \right.  \left. \left| \Psi_{J_iM_i}
 (1\cdots A) \right\rangle \right\}\ ,
\end{eqnarray}
where $\Omega_{sc}$ is the scattering angle and  ${\cal A}_{01}$ is 
the antisymmetrization operator. 
The nuclear transition is from a state $\vert J_i M_i \rangle$
to a state $\vert J_f M_f \rangle$ and the projectile has spin projections
$\nu$ and $\nu^\prime$ before and after the collision with the incoming
and outgoing distorted waves being $\chi^{\pm}$.
The incoming and outgoing relative momenta are ${\bf k}_i$ and  ${\bf k}_o$
respectively.  Then a cofactor expansion of the target states, 
\begin{equation}
\left| \Psi_{JM}(1\cdots A) \right\rangle = \frac{1}{\sqrt{A}} \sum_{j,m}
\left| \varphi_{jm}(1) \right\rangle\,
a_{jm}(1)\, \left| \Psi_{JM}(1\cdots A) \right\rangle\ ,
\label{cofactor}
\end{equation}
allows  expansion of the  scattering  amplitudes in the form  of 
weighted two-nucleon elements since the terms 
$a_{jm}(1)\, \left| \Psi_{JM}(1\cdots A)\right\rangle$ in 
Eq.~(\ref{cofactor}) are independent of coordinate `1'. Thus 
\begin{eqnarray}
{\cal T} &=& \sum_{j_1,j_2} \left\langle\Psi_{J_fM_f}(1\cdots A) \right|
a^{\dagger}_{j_2m_2}(1) a_{j_1m_1}(1) \left| \Psi_{J_iM_i}(1\cdots A) 
\right\rangle 
\nonumber\\
&&\hspace*{1.0cm} \times
\left\langle \chi^{(-)}_{\nu^\prime}({\bf k}_o0)\right|
\left\langle \varphi_{j_2m_2}(1) \right| \ {\bf g}_{\text{eff}}(0,1)\ 
{\cal A}_{01} \left\{ \left| \chi^{(+)}_\nu ({\bf k}_i0) 
\right\rangle
\ \left| \varphi_{j_1m_1} (1) \right\rangle \right\}
\nonumber\\
&=& \sum_{j_1,j_2,m_1,m_2,I(N)}
(-1)^{(j_1-m_1)} \frac{1}{\sqrt{2J_f+1}}\,  
\left< J_i\, I\, M_i\, N \vert J_f\, M_f \right>
\left< j_1\, j_2\, m_1\, -m_2 \vert I\, -N \right>
\, S_{j_1\, j_2\, I}^{(J_i \to J_f)}\ 
\nonumber\\
&&\hspace*{2.0cm} \times
\Bigl< \chi^{(-)}_{\nu^\prime}({\bf k}_o0)\Bigr|
\left\langle \varphi_{j_2m_2}(1) \right| \
{\bf g}_{\text{eff}}(0,1)\ 
 {\cal A}_{01} \left\{ \left| \chi^{(+)}_\nu ({\bf k}_i0) 
\right\rangle
\ \left| \varphi_{j_1m_1} (1) \right\rangle \right\}\ ,
\end{eqnarray}
where reduction of the structure factor to  OBDME  for    angular 
momentum transfer values $I$ follows that  developed  earlier~\cite{Am05}.

  The effective interactions $g_{\text{eff}}(0,1)$ used in the folding to
get  the  optical  potentials   have   also   been   used  as  the  
transition  operators  effecting  the excitations    (of the $6^-$ 
states). As with the generation of the elastic scattering optical
potentials from which the distorted waves are generated,
antisymmetry   of   the  projectile  with   the  individual  bound 
nucleons is treated exactly.   The associated knock-out (exchange) 
amplitudes   contribute   importantly   to  the  scattering  cross  
section, both in magnitude and shape~\cite{Am00}.
The OBDME and single particle wave functions required in these matrix 
elements are those specified in the previous section.

\section{Results}
\label{Results}

\subsection{Elastic scattering of protons from ${}^{28}$Si}
\begin{figure}[h]
\scalebox{0.7}{\includegraphics*{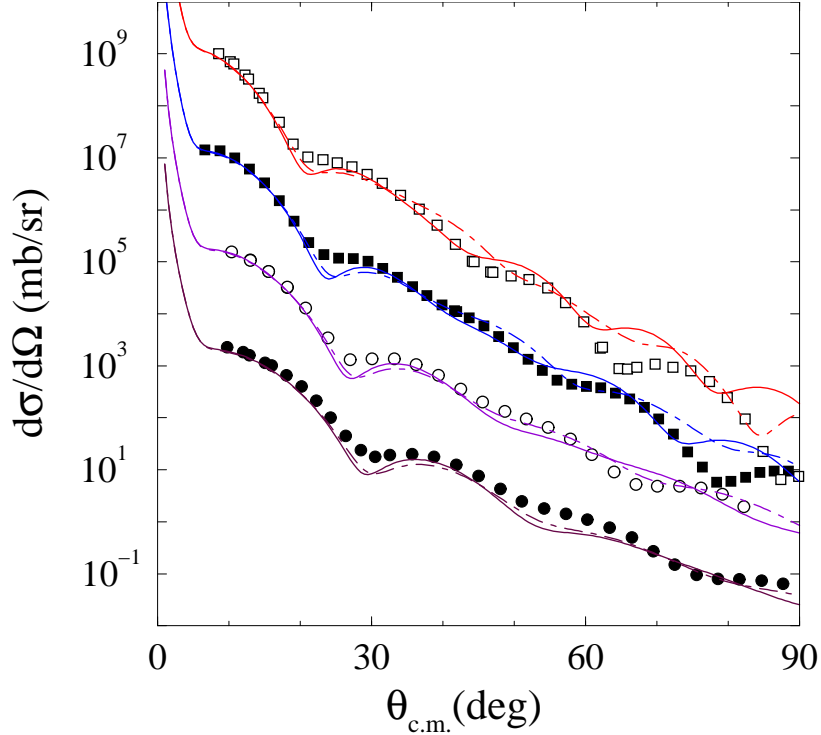}}
\caption{\label{Fig3}(Color online)
$g$-folding model predictions of  cross sections for the elastic scattering 
of 80, 100, 134, and 180 MeV protons from ${}^{28}$Si compared with data.
The predictions found by using the SHF structure are shown by the solid
curves while those obtained with the WS single particle wave functions
are displayed by the dash-dot curves.}
\end{figure}

The cross sections for the elastic scattering of 80, 100, 134, and 180~MeV
protons from ${}^{28}$Si calculated using the $g$-folding model for the
relevant optical potentials are compared with data~\cite{Ol84} in 
Fig.~\ref{Fig3}.
Two potentials for each energy were formed, one set by folding the effective
$NN$ interactions with the SHF wave functions and the other by using WS
single particle wave functions. In all cases, the orbit occupancies
were those specified previously in Table~\ref{Occup}.
The 80 MeV data and results are unscaled and are the lowest shown in the 
figure. For clarity, the other three results (and data) have been multiplied 
by 10$^2$, 10$^4$, and 10$^6$ for the 100, 134, and 180 MeV cases 
respectively. Our results, while good in comparison 
with the data, are not as perfect fits as has been found~\cite{Ol84}.
But that study~\cite{Ol84} used a phenomenological potential whose 
parameter values were determined by numerical inversion of data.
In contrast, our results are predictions
resulting from the use of model structures folded over effective interactions
and with no {\it a posteriori} adjustments of any kind.
The solid curves in this figure are the results obtained using the SHF 
model of structure while the dot-dashed curves are those obtained when WS bound 
state wave functions are used in place of the canonical ones of the SHF.
The comparisons with data are quite good for cross section values larger 
than $\sim 10^{-1}$ mb/sr with the SHF predictions 
being preferable in general. Nevertheless there are some disagreements between
the calculated and the experimental cross sections which, in view of the 
variations found between the SHF and the WS cross sections, bespeaks of
a needed improvement in the details of the structure model. 
Particularly it is the mismatch to data having small values at larger
scattering  angles (and so larger momentum transfer values) that needs
addressing in future. These disparities indicate that the structure
of the densities inside  the nucleus are most likely to blame.   

\begin{figure}[h]
\scalebox{0.7}{\includegraphics*{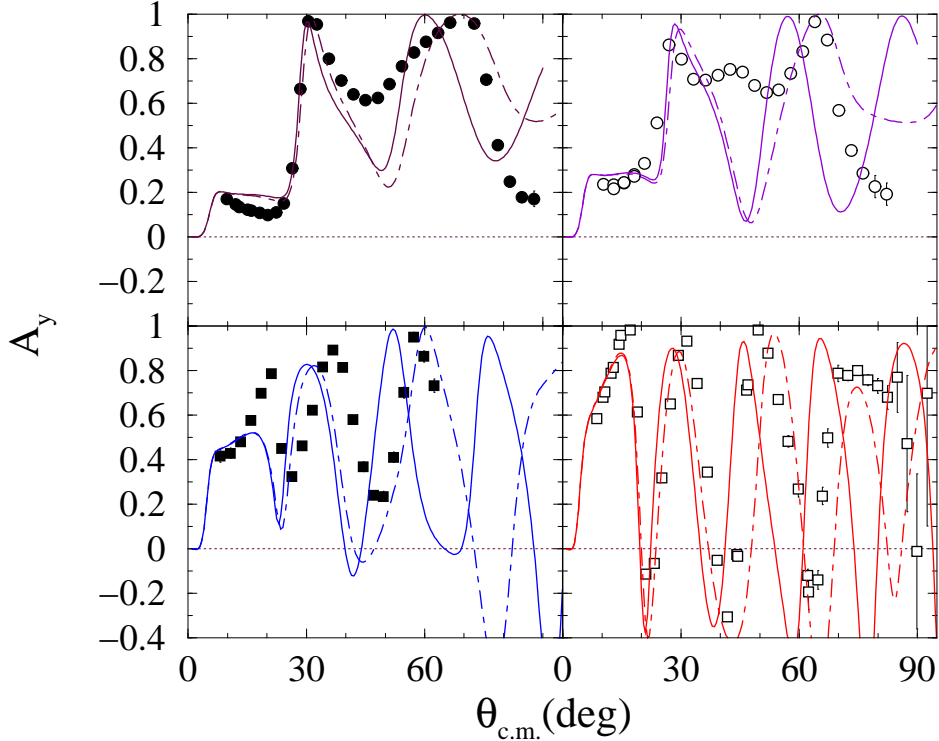}}
\caption{\label{Fig4}(Color online)
$g$-folding model predictions of analyzing powers  for the elastic scattering 
of 80, 100, 134, and 180 MeV protons from ${}^{28}$Si compared with data.
The notation is as used in Fig.~\ref{Fig3}.}
\end{figure}

That is emphasized
in the comparison between our predictions and the associated analyzing 
power data~\cite{Ol84}.
The elastic scattering  analyzing power data and $g$-folding model 
predictions for the four energies considered are displayed in Fig.~\ref{Fig4}.
The notation is the same as used in Fig.~\ref{Fig3} with the 80 MeV results 
shown in the top left panel, the 100 MeV results placed in the top right panel
and the 134 and 180 MeV results presented in the bottom, left and right panels, 
respectively.
There is obvious mismatch between predictions and data of this quantity.
Also there is greater variation between the predictions made with the SHF 
structure and those made with the WS functions. Nonetheless the general trend 
of the data is found, more so with the 134 and 180 MeV results. 
Again we note that the phenomenological approach does far better in fitting 
than this but it must be remembered that in the phenomenological process,
parameter adjustments are designed to
yield best fits. That is so even if any data may be suspect. When using
phenomenology, some independent condition or conditions (besides a fit to data)
should be applied in defining a `good' phenomenological potential.
The analyzing power results are much more sensitive to the nature of the
single particle bound state wave functions used to form the $g$-folding
potentials. This was noted also in the review~\cite{Am00} for many
target masses, and at energies of 65 and 200 MeV notably. We have found 
smooth variations in these observables from 80 to 180
MeV; much smoother than the existent data in fact. However, all we may claim
at this stage is that improved wave functions with which better fits to
the cross section data need be found before
more concrete statements resulting from the comparisons of analyzing powers
can be made.

\subsection{Inelastic proton scattering to $6^-$ states of pure isospin}

\begin{figure}[h]
\scalebox{0.7}{\includegraphics*{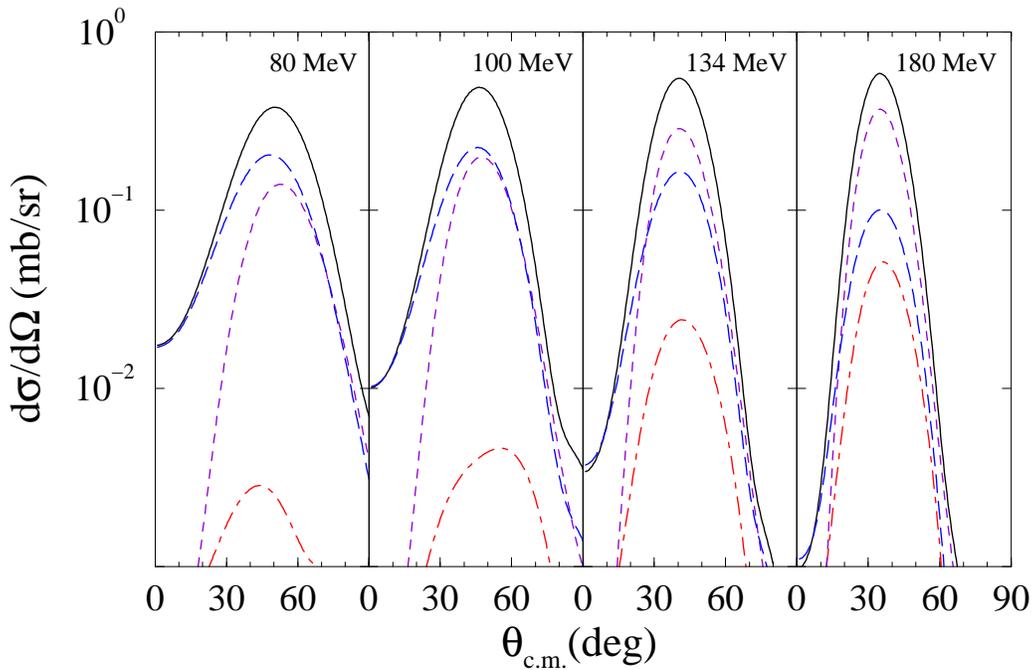}}
\caption{\label{Fig5}(Color online)
DWA cross sections for the inelastic scattering of 80, 100, 134,
and 180 MeV protons from ${}^{28}$Si exciting a 6$^-$ (T=0) state.
In each panel the solid curves are the full calculated results, 
the long dashed curves depict those generated by purely the two-body
tensor force as the transition operator, the small dashed curves are 
those generated by the two-body spin-orbit force alone, and the 
dot-dashed curves depict those cross sections generated using purely
 the central force components.}
\end{figure}

Cross sections for the excitation of an 
isoscalar 6$^-$ state in ${}^{28}$Si at 11.58 MeV excitation resulting
from DWA calculations are displayed in Fig.~\ref{Fig5}.  The structure used
in those calculations was that of a pure particle-hole excitation as 
detailed above based upon the SHF ground state. 
Therein, as in the ensuing figures, the results for 80, 100, 134, and 
180 MeV are displayed in panels from left to right (as indicated). 
The full result (solid curve) is compared to those found when each of the 
three components of the effective transition operator are used by themselves,
namely the central (dot-dash curve), the two-body spin-orbit 
$\mathbf{L} \cdot \mathbf{S}$ (dash curve), and the two-body tensor 
$\mathbf{S}_{12}$
(long dash curve) components. For this pure isoscalar transition, the central 
force component of the transition operator contributes only in a minor way
gradually being more influential with increasing energy. In contrast, the 
contribution from the tensor force is substantial but slowly decreases with 
increasing energy. The two-body spin-orbit contributions are also substantial
and increase to be the dominant feature of the scattering  at 180 MeV. 
The results and the relative contributions are similar to those found 
previously~\cite{Ol84} with the Love-Franey effective interaction. However,
not only were those calculations~\cite{Ol84} made using a distorted
wave impulse approximation (DWIA) and hence did not evaluate explicit
exchange matrix elements due to the Pauli principle, but also the Love-Franey 
force was one formed under a constraint to fit a set of free $NN$ scattering 
data and did not allow for medium modifications. 
Additionally, those older calculations also used phenomenological, local,
optical potentials to generate the distorted waves with the parameters of 
those potentials chosen to find good fits to elastic scattering data.
But that only requires specification of a suitable set of phase shifts and
they are determined from the asymptotic forms of the distorted waves.
The credibility of the distorted wave functions through the 
volume of the nucleus, properties needed in evaluation of inelastic scattering 
amplitudes, cannot be ascertained.

\begin{figure}[h]
\scalebox{0.7}{\includegraphics*{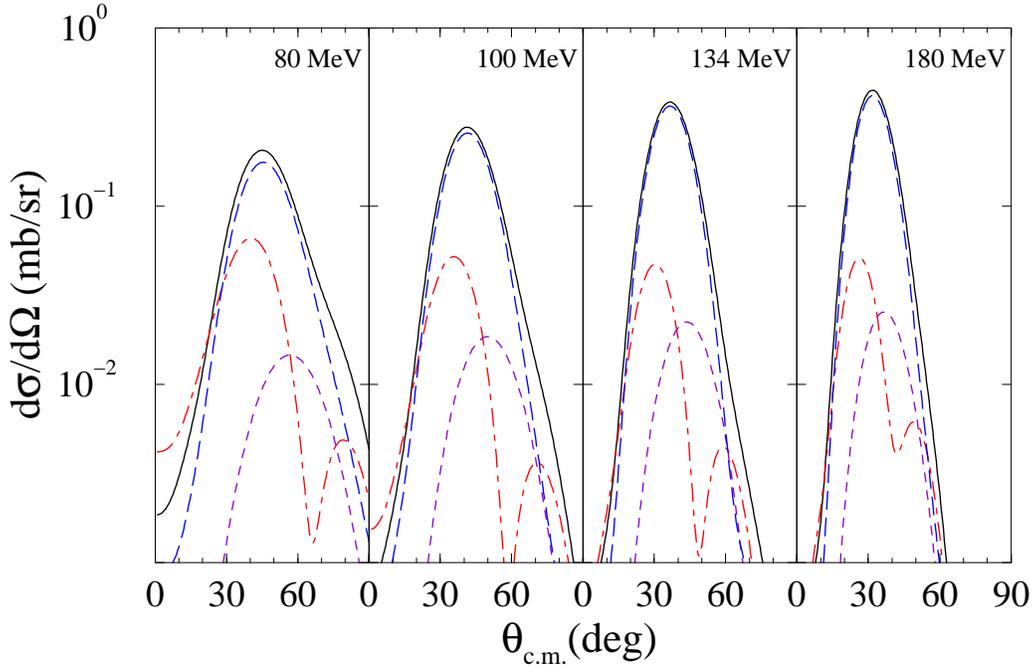}}
\caption{\label{Fig6}(Color online)
DWA cross sections for the inelastic scattering of 80, 100, 134,
and 180 MeV protons from ${}^{28}$Si exciting a 6$^-$ (T=1) state.
The notation is as used in Fig.~\ref{Fig5}.}
\end{figure}

In Fig.~\ref{Fig6} we show the cross sections obtained from DWA 
calculations of inelastic proton scattering at four energies and
exciting a 6$^-$ (T = 1) state at 14.35 MeV. The notation is as used in 
Fig.~\ref{Fig5}.
For this pure isovector transition, the tensor force is the dominant
contributing component of the transition operator. It increases in 
magnitude with increasing energy so that at all four energies, the 
contributions from  the central and two-body spin-orbit components
remain minor. However they have distinctively different cross section
component shapes with the effect of the central force 
being evident in the total result at forward angles and at 80 MeV.

\begin{figure}[h]
\scalebox{0.6}{\includegraphics*{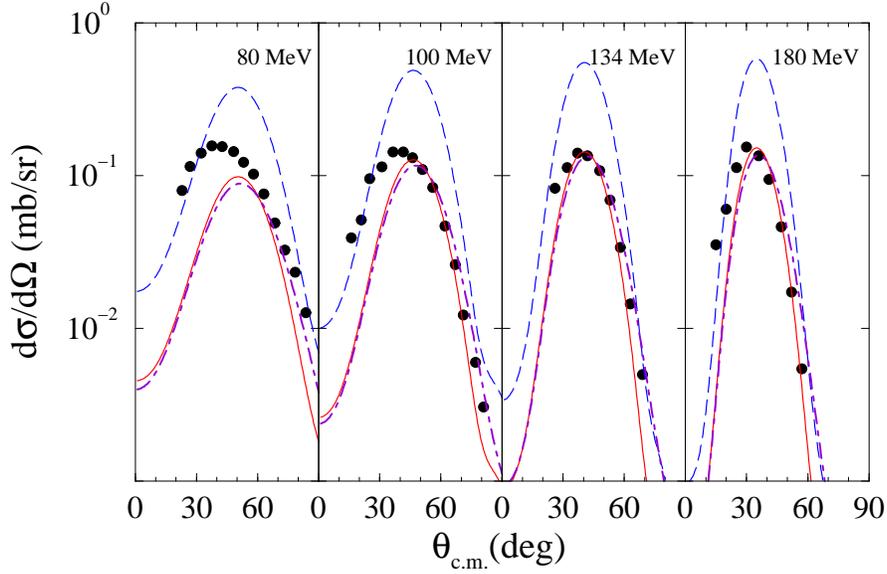}}
\caption{\label{Fig7}(Color online)
Cross sections for the inelastic scattering of 80, 100, 134,
and 180 MeV protons from ${}^{28}$Si exciting a 6$^-$ (T=0) state
compared with data~\cite{Ol84}.} 
\end{figure}

Fig.~\ref{Fig7} displays the cross sections from the excitation of the 
11.58
MeV state treated as a pure isoscalar transition. The cross sections found
by using the SHF wave functions and the OBDME for the simple particle-hole
description of the excited state are depicted by the dashed curves. At all 
incident energies, the data are overestimated by those results. 
The solid curves then are
those same cross sections multiplied by a factor of 0.26. For comparison,
the cross sections found by using the WS wave functions, and with a similar 
scale reduction, are depicted by the dot-dashed curves. The choice of
wave functions does not make significant change in the cross section 
results.
The shapes and magnitudes of the 134 and 180 MeV data are quite well 
reproduced by these scaled, pure isoscalar, calculated cross sections.
The 100 MeV scaled cross section replicates the overall magnitude and 
the large angle data reasonably but there is a distinct disparity between
result and data at the forward scattering angles. Those disparities are 
even more evident in the 80 MeV results. Such were the case also in the 
previous analyses~\cite{Ol84}.
\begin{figure}[h]
\scalebox{0.6}{\includegraphics*{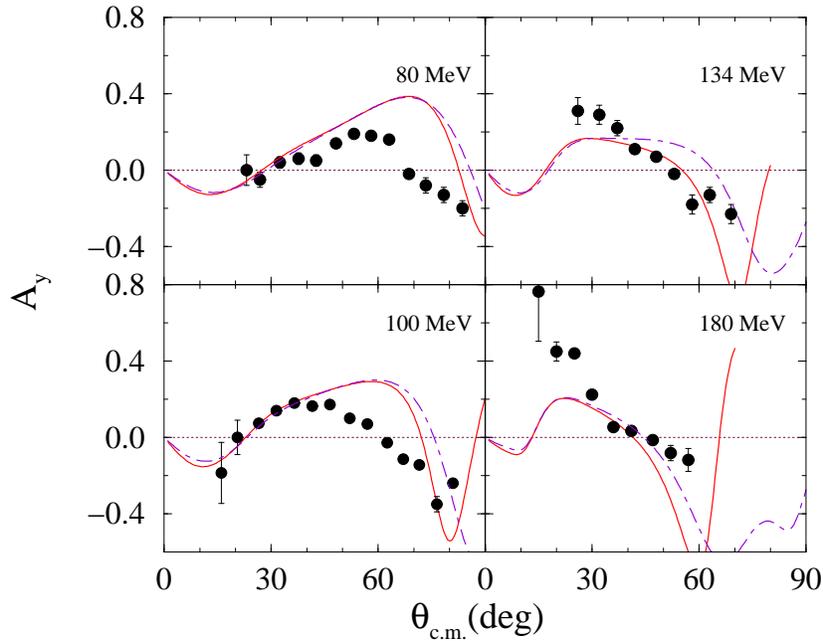}}
\caption{\label{Fig8}(Color online)
Analyzing powers for the inelastic scattering of 80, 100, 134,
and 180 MeV protons from ${}^{28}$Si exciting a 6$^-$ (T=0) state
compared with data~\cite{Ol84}.}
\end{figure}

The analyzing powers from this isoscalar transition are depicted 
in Fig.~\ref{Fig8} with the SHF and WS results again being displayed by
the solid and dot-dashed curves respectively. The agreement between 
our calculated results and
the data is only average in quality, but the trend of the data is
replicated. As energy increases,, the positive peak in the data 
moves to smaller scattering angles and the data give negative values
at the larger scattering angles. So also do our calculated results.
In this observable the choice of wave functions (SHF or WS) has a slightly 
more noticeable effect than with the cross sections, particularly
at large angles for the higher energies.

\begin{figure}[h]
\scalebox{0.6}{\includegraphics*{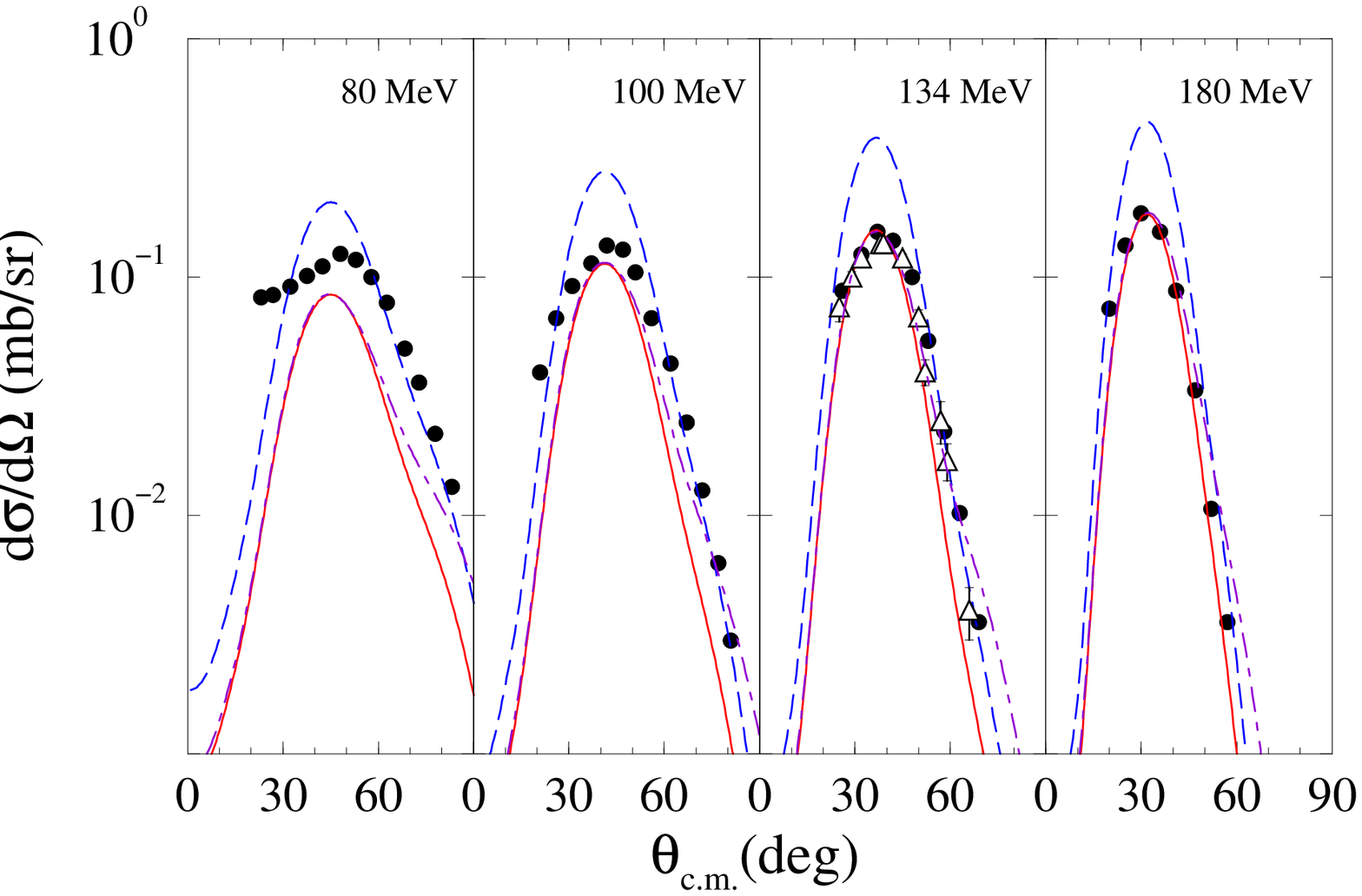}}
\caption{\label{Fig9}(Color online)
Cross sections for the inelastic scattering of 80, 100, 134,
and 180 MeV protons from ${}^{28}$Si exciting a 6$^-$ (T=1) state
compared with data~\cite{Ol84}. The 134 MeV ($p,n$) data~\cite{Fa85} 
($\times \frac{1}{2}$) exciting the analogue state in ${}^{28}$P 
are shown by the open triangles. The full calculation results are
displayed by the dashed curves while the solid curves result
when a scale of 0.41 is applied.} 
\end{figure}
\begin{figure}[h]
\scalebox{0.6}{\includegraphics*{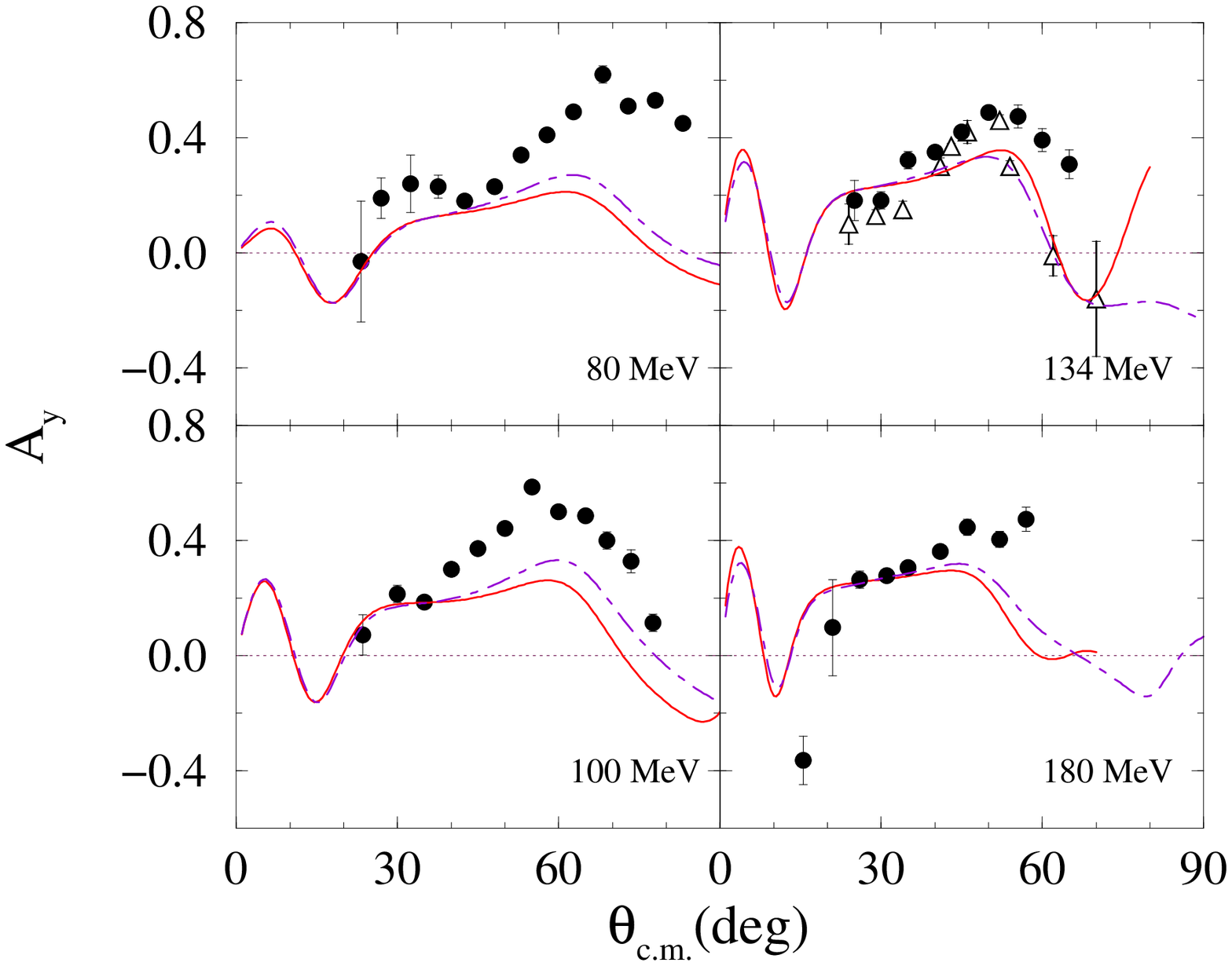}}
\caption{\label{Fig10}(Color online)
Analyzing powers for the inelastic scattering of 80, 100, 134,
and 180 MeV protons from ${}^{28}$Si exciting a 6$^-$ (T=1) state
compared with data~\cite{Ol84}. The 134 MeV ($p,n$) data~\cite{Fa85}
exciting the analogue state in ${}^{28}$P are shown by the open triangles.
The solid curves depict the results found using the SHF wave functions
while the dot-dashed curves are those found using the WS functions.} 
\end{figure}

In Fig.~\ref{Fig9} cross-section data~\cite{Ol84} from
inelastic proton scattering to the 14.35 MeV state of ${}^{28}$Si are 
compared with our DWA results considering the transition to be purely 
isovector. In this case the 
solid curves depict results found using a scale factor of 0.41 on
the total cross sections. While a slightly smaller value (0.36) was 
required to match the electron scattering form factor, the results are
sufficiently similar to consider that the two data sets confirm the
degree of configuration mixing in the isovector $6^-$ state. Again the 
results obtained by using the WS 
instead of the SHF wave functions are depicted by the dot-dashed curves.
For this 14.35 MeV excitation, we note that 
the comparisons between data and calculated results are very similar to 
those found with the isoscalar transition. However,
the disparities seen with energies of  80 and 100 MeV are accentuated
while the 134 and 180 MeV results are somewhat in better agreement with data.
In this figure we also include the cross sections measured at 134 
MeV~\cite{Fa85}
from the charge exchange ($p,n$) reaction to the analogue $6^-$ state
in  ${}^{28}$P. The actual data has been halved in accord with the isospin 
weighting
coefficient that relates the ($p,n$) scattering to the ($p,p'$) scattering 
to the $6^-; T=1$ analogue state in ${}^{28}$Si. 
The ($p,n$) data ($\times \frac{1}{2}$) are portrayed by the open triangles.
That the two cross sections
agree so well confirms the analogue character of the states involved and also
argues for purity in isospin. 

The analyzing powers associated with the cross sections discussed immediately
above, are compared with data in Fig~\ref{Fig10}. 
Again there are some mismatching to the 80 and 100 MeV data
though the trend as the incident energy increases is reflected by
the calculated results. Now, however, the 134 and 180 MeV data are quite well 
reproduced, especially when one considers the ($p,n$) reaction values at
134 MeV.

We have made numerous evaluations also allowing degrees of isospin mixing
in the two states for which there are data. That such could occur follows 
from the strong indication we found~\cite{Am05} for such a mixing in the 
$4^-$ states in the spectrum of ${}^{16}$O. In this case though the two 
$6^-$ states to which there are scattering data lie $\sim 3$ MeV apart, there
are a number of other $6^-$ states around and between them. However, no 
improvement in the shape of cross sections compared to the data results
for any reasonable admixing of isospin, nor is there any significant 
improvement in the peak magnitude variation with energy. 

\section{Conclusions}
\label{Conclusions}

We have analyzed proton scattering cross sections and analyzing powers
from the scattering of protons at 80, 100, 134, and 180 MeV from
${}^{28}$Si. Both elastic scattering data and those from the excitation of
two $6^-$ states have been considered. Those states, thought to be stretched, 
single particle-hole excitations upon the ground state, are excited dominantly
by the spin-orbit and tensor force terms in the effective $NN$ interaction
that is considered as the transition operator.

The 134 and 180 MeV data, elastic and inelastic, cross sections and analyzing 
powers, are quite well reproduced by our predictions when some configuration
mixing is allowed to fragment the particle-hole strength among other known
$6^-$ states in ${}^{28}$Si.
The complementary analyses of the M6 electron scattering form factor and of
the $(p,p^\prime)$ observables at 134 and 180 MeV exciting the (isovector)
14.35 MeV state strongly indicate that this transition exhausts $\sim 60$\%
of the isovector particle-hole strength. There are no electron scattering data
from excitation of the 11.58 MeV state from which to extract a form factor,
 but the analyses of the 134 and 180 MeV 
$(p,p^\prime)$ cross sections suggest that this transition is isoscalar 
and exhausts $\sim 50$\% of that particle-hole strength.

The data taken with  80 and 100 MeV protons is not as well reproduced, notably
the elastic scattering analyzing powers and the peak magnitudes for the $6^-$
excitations found using the same scale required to match data at 134 and 
180 MeV. We note that the shape of the elastic scattering data do not reflect
the smooth trend with energy at large scattering angles  
given by our microscopic model calculations. With the inelastic scattering 
cross sections, the 80 MeV, and 
to a lesser extent the 100 MeV, data also are larger at forward scattering 
angles than our predictions. 
Of course, these inelastic excitations are particularly sensitive to the
spin-orbit and/or the tensor character of the effective $NN$ interaction.
While the results at 134 and 180 MeV, taken in conjunction with those
for excitation of the $4^-$ states in ${}^{16}$O with 134 MeV 
protons~\cite{Am05}, do give credibility to the relevant components of
the  effective $NN$ interaction at those energies, on the basis of the data 
considered, the 
strengths of the tensor components at 80 and 100 MeV may be weak.  
Analyses of more data
from scattering with proton energies at and below 100 MeV,
and from other stretched state excitations, are needed to ascertain 
any such adjustment to the effective interaction. 

\begin{acknowledgments}
This research was supported  by a research grant from the
Australian Research Council and by a grant from the Cheju National
University Development Foundation (2004).
\end{acknowledgments}

\bibliography{Si28-6m}

\end{document}